\documentclass[12pt,letterpaper,onecolumn]{article}
\usepackage[latin1]{inputenc}
\usepackage{amsmath}
\usepackage{amsfonts}
\usepackage{amssymb}
\usepackage{braket}
\usepackage[left=1.00in, right=1.00in, top=1.00in, bottom=1.00in]{geometry}
\usepackage{enumerate}
\usepackage{graphicx}
\author{Horace P. Yuen\\
Department of Electrical Engineering and Computer Science\\
Department of Physics and Astronomy\\
Northwestern University, Evanston Il. 60208\\
yuen@eecs.northwestern.edu
}
\title{FUNDAMENTAL AND PRACTICAL PROBLEMS OF QKD SECURITY-THE ACTUAL AND THE PERCEIVED SITUATION\footnote{A previous version of this paper is for the Proceedings of the SPIE meeting on Security and Defence in Prague, September, 2011.}}
\begin{document}
\maketitle

\newpage\begin{abstract}
\textit{It is widely believed that quantum key distribution (QKD) has been
proved unconditionally secure for realistic models applicable to various
current experimental schemes. Here we summarize briefly why this is
not the case, from both the viewpoints of fundamental quantitative
security and applicable models of security analysis, with some morals
drawn.}
\end{abstract}

\section{Introduction}

After the appearance of papers last year on fundamental QKD security
{[}1{]} and the complete breach of concrete QKD systems {[}2-3{]},
claims have persistently been made {[}4{]} that QKD is already proved
unconditionally secure in principle in various models. However, ref
{[}2{]} and its extensions {[}3,5{]} highlight in a forceful manner
the precarious situation of such widespread claims, especially on
concrete experimental systems. Imagine the consequence if the Norway
group kept their detector ``blinding attacks''
secret and makes them available to selected parties after a QKD systems
has been deployed upon convincing the users its unconditional
or whatever imposing security terminology that has been employed.
The fact of the matter is that the security proofs of the models,
assuming the deductions are totally valid (but they are actually not),
contain general and specific assumptions that are simply not satisfied
in practice. Moreover, the security criteria themselves used in the
proofs do not guarantee proper security when satisfied. This paper
tries to outline the underlying reasons, and indicates some ways to
deal with the situation.

\section{Security is not merely a matter of definition}

Thus far a single-number criterion on a single quantity has been used
as the security criterion in QKD security proofs, from mutual information
to trace distance. However, in accord with detection theory an attacker
Eve would obtain from the measurement result on her probe a whole
probability distribution $\{q_{j}\},\; j\in{1,\ldots,M}$ for the
$M=2^{m}$ possible values of an m-bit data $X$ from which an n-bit
final key $K$ is drawn after the use of an error correction code (ECC) and a privacy
amplification code (PAC). Since PAC is a known hash function, this
distribution $\{q_{j}\}$ $X$ leads to the distribution $\{p_{i}\}$
on $K$, $i\in\{1,\ldots,N\}$. A bound on the mutual information or
any single-number criterion on $X$ or $K$ merely expresses a constrain
on $\{q_{j}\}$ or $\{p_{j}\}$.

Since one is using $K$ as if it is uniformly
distributed to Eve, i.e., with probability $\frac{1}{N}$ for the
N possible key values, one must bound the difference between the probability
$2^{|\tilde{K}|}$ of a uniformly distributed subset of size $|\tilde{K}|$
and Eve's optimum probability $p_{1}^{E}(\tilde{K})$ of estimating
$\tilde{K}$ correctly in any attack to be below a prescribed security
level,

\begin{equation}
|p_1^E(\tilde{K})-2^{-|\tilde{K}|}|\leq\epsilon(\tilde{K}) \;\;\;\; \text{for each} \;\;\;\; \tilde{K}\subseteq K
\end{equation}

When $\varepsilon(\tilde{K})$ can be made sufficiently small, condition
(1) gives $K$ the information-theoretic (IT) ``
security'' required for meaningful ``unconditional
security''. Thus, good security demands semantic security
that may not be obtainable quantitatively from another criterion,
especially a single-number criterion that is not bounded tightly enough.
Security is a quantitative issue. It is also not a mere mathematical
issue for which one can adopt whatever mathematical definition that
seems intuitively suitable. In particular, it must be expressed by
Eve's success probabilities of correctly estimating various properties
of $K$ which can all be derived from (1).

Eve can try to estimate $K$ from just its generation process, the resulting
security of $K$ is called ``raw security''
to be distinguished from its ``composition security''
when $K$ is actually used. General composition security is a complicated
matter and we will just restrict our attention to the case when $K$
is used for encryption, say in the one-time pad format often suggested.
In such situation part of $K$ may be revealed to Eve in a known-plaintext
attack (KPA), which may help her find the rest of $K$ and thus find
the rest of the data unknown to Eve at the beginning, which is encrypted
by $K$ to yield the ciphertext. Security against such leakage of $K$ will
be called ``KPA security''.

It is important to note that proof of general KPA security is necessary
for any claim of unconditional security on $K$ to be used for encryption.
This is because the use of conventional symmetric key ciphers for
key expansion also gives raw IT security,
and thus they, not RSA, are the appropriate conventional ciphers one
should compare QKD to. This is a fair comparison because the use of
a shared secret key for message authentication during key generation
is necessary, though not included thus far in the security analysis
of any QKD protocol. (KCQ protocols {[}6{]} use a shared secret key
explicitly.) In addition, the raw security of such conventional ciphers
is far better than that of concrete QKD systems that have been studied
experimentally or theoretically. The superiority of concrete QKD must
lie in its KPA security, which is the usual security concern because
the shared secret key is typically totally hidden when the data $X$
is uniform to Eve.

\section{Problems of the mutual information criterion}

Eve's accessible information on $K$ from her attack is the most commonly
used security criterion, so far the only one used in all experimental
schemes. It is Eve's mutual information $I_{E}$ with respect to $K$
under optimal measurement on her probe. Information or entropy expresses
a constraint on Eve's estimate on the whole distribution $\{p_{j}\}$
she may get from the measurement result on her probe. It has been
repeatedly pointed out {[}7,8,1{]} that there are distributions consistent
with a given $I_{E}$ such that her maximum probability $p_{1}$ of
estimating the whole key $K$ correctly is given by 
\begin{equation}
p_{1}\sim\frac{{I_{E}}}{n}\equiv2^{-l}
\end{equation}
Thus, unless $l\sim n$, the raw security of $K$ so guaranteed
may be quite inferior to a uniform key. The other subsets $\tilde{K}$
of $K$ suffer similarly. The practical values of $I_{E}$ obtained in
experimental schemes indeed gives very large $p_{1}$ in this sense
{[}1,8{]}.

Under KPA, knowing some bits of $K$ does not render the rest of $K$ more
insecure if E has no quantum memory {[}1{]}. If Eve does have quantum
memory, possible locking information would render $K$ insecure {[}9{]}
or even very insecure {[}10{]}. In fact, the latter can be understood
from (2) as follows. The bits on $K$ gained in a KPA could reduce the
exponent of $p_{1}$ in (2). Indeed, it only takes 
\begin{equation}
l^{'}=l+\log n
\end{equation}
number of bits to change $p_{1}$ to the value 1 when Eve
measures on her probe with this added information on $K$.

The variational distance
\begin{equation}
\delta_{E}\equiv\delta(P,U)=\frac{{1}}{2}\sum_{i}|p_{i}-U|
\end{equation}
between Eve's probability distribution $\{p_{i}\}$ on $K$ averaged
over $K$ and the uniform distribution $U$ of n bits
has quantitative behavior similar to $I_{E}$. In particular, when
$p_{1}$ is large compared to $\frac{1}{N}$,$\delta_{E}$ could give
a $p_{1}$ as big as the case of $\delta_{E}=\frac{I_{E}}{N}$, i.e.
the value of (2). Thus, the same problem occurs under KPA as in the
case of an $I_{E}$ criterion above. When $\delta_{E}\le\varepsilon=2^{-l}$
with $l\sim n$ , good security close to $U$ may be obtained.

\section{Problems of the trace distance criterion}
The trace distance quantum criterion
\begin{equation}
\frac{{1}}{2}\parallel\rho_{1}-\rho_{2}\parallel_{1}\le\varepsilon
\end{equation}
between two density operators $\rho_{1}$ and $\rho_{2}$ says that
the variational distance between the two distributions $P$ and $Q$
obtained in a measurement as derived from $\rho_{1}$ and $\rho_{2}$
satisfies $\delta(P,Q)\le\varepsilon$. Let $\rho_{E}^{k}$ be the
state of Eve's probe when the actual $K$ has value $k$. Then (5) says,
with $\rho_{1}=\rho_{E}^{k}$ and $\rho_{2}=\rho_{U}$ the uniform
mixed state with rank N, that for any measurement Eve may make on
her probe one has $\delta_{E}\le\varepsilon$. The problem of such
a security criterion is indicated through $\delta_{E}$ above.

With $\rho_{E}=E_{k}[\rho_{E}^{k}]$, the following trace distance
\begin{equation}
d=\frac{{1}}{2}\parallel\rho_{KE}-\rho_{U}\otimes\rho_{E}\parallel_{1}
\end{equation}
for a joint state $\rho_{KE}$, is used [11] with the interpretation
that when $d=\epsilon$, it means $K$ equals $U$ with probability $1-\varepsilon$
to Eve and the value k also becomes independent of $\rho_{E}^{k}$.
This implies ``universal composability''
including security against KPA. In ref [14] it has been analyzed
in detail why this interpretation cannot be true with any probability.
We can describe the reason simply as follows.

The main error arises from conclusion in ref {[}11{]} that $\delta(P,Q)=\epsilon$
implies $P$ and $Q$ are the same distribution with probability $1-\varepsilon$.
This conclusion was derived from the existence of a joint distribution
$D$ that gives $P$ and $Q$ as marginal and yields the above interpretation.
Why would this $D$ arise in the cryptosystem? In fact, even when
a joint distribution different from the product form $PQ$ is in force,
why would it be this particular (which is actually the optimal) one
for the interpretation to obtain. In reality, one is simply comparing
two distributions and the joint distribution should be $PQ$. It is
indeed clear directly from the definition of $\delta(P,Q)$ that $P$
and $Q$ must differ when $\delta(P,Q)=\varepsilon>0$.

Sometimes the term \textquotedblleft{}failure probability\textquotedblright{}
is used {[}12,13{]} without explicitly saying what that means. In
{[}14{]} it is shown that $\varepsilon$ does not itself have a probability
interpretation. Thus, it is not any {}``failure probability''.
A related source of this error, which was not discussed in ref {[}1{]},
is a misinterpretation of a notion of {}``$\varepsilon$-indistinguishable''
measure. It is concluded [12] from (5) that $\rho_{1}$ and $\rho_{2}$
is ``$\epsilon$-indistinguishable''
and thus the protocol has \textquotedblleft{}failure probability\textquotedblright{}
$\ge1-\varepsilon$. The problem of {}``$\varepsilon$-indistinguishable''
for KPA security guarantee is a quantitative one similar to (2)-(3)
above. A detailed explanation of the whole situation is given in [15].

\section{True key generation rate and limitation of privacy amplification}
It is important to observe that Eve\textquoteright{}s maximum probability
$q_{1}$ on the data $X$ is equal to her $p_{1}$ on $K$, and that $p_{1}$
\textit{cannot} be be improved by further PA. This is because a known
transformation from ECC+PAC would just bring the most likely value
of the data or key to a final value of $K$ with the same probability.
On the other hand, it is clear from (2) that the rate of secure key generation is limited to $\frac{l}{n}$. Thus, unconditionally
secure key generation rate \textit{cannot} be given by what has been
asserted in the literature, but is determined via the $p_{1}$ exponent.
Moreover, this rate is determined by $q_{1}$ and we just saw it \textit{cannot}
be improved via PA. 

In finite protocols -- no real protocol operates in the asymptotic
limit $n\rightarrow\infty$ -- it makes little sense to say a quantity
grows exponential in $n$ without some estimate of the actual convergence
rate, because any value can be written as exponential in $n$. It
is more accurate to just say that $l$ secure bits are
generated in the round. For concrete protocols $l$ is very small
thus far and may not even cover the message authentication bits used
in any normal IT-secure authentication scheme. 

\section{True security and asymptotic proof}

Note that even with PA to extract semantically secure bits from the
$p_{1}$ exponent, fundamental security has not been guaranteed. The
use of Markov Inequality to convert an average guarantee to a probable
individual guarantee would just reduce the quantitative value that
has been achieved. Much more significantly, Eve could launch an optimal
quantum attack on specific subset $\tilde{K}$ of $K$ which, because
of quantum mechanics, can be superior to what she may obtain by attacking the whole $K$
and must be bounded in a security proof.

In this connection, it is important to point out that the asymptotic
limit that so many security proofs are based upon both overestimate
what the users can achieve and what Eve can achieve in a finite situation.
This is clear information theoretically, because all the ``capacity''
like statements involving mutual information are well known to be
limiting capabilities. For example, the actual ``random
coding exponent'' or channel reliability function {[}16{]}
for finite $n$, similar to the $p_{1}$, exponent, is what controls
finite system performance, not the capacities. This applies to both
the users and Eve, and is overlooked also in the classical literature
on key generation. Basically, cryptographers should be working with
detection theory, not information theory, for ascertaining performance
by any party. Probability has operational interpretation and is what
matters IT wise (but IT in the broad sense), not any other theoretical
quantity like mutual information that needs to be translated back
to probability as done in ref {[}1,7{]}.

This last point is very important. It shows the possibility of secure
key generation is not determined by any capacity statement. Indeed,
in KCQ (keyed communication in quantum noise) {[}7{]} one does not allow 
coding or indefinitely large $n$ for Eve other than her optimum decision 
on a finite-n system. Quantum information
locking may help significantly for KCQ but it is not necessary.

Some details and further elaboration on sections II-VI can be found
in {[}14,15{]}.

\section{Grave effect of loss on security}

Real optical systems have significant loss. If the transmission loss
is small one can treat deleted bits as random errors. Security claim
was often made with loss taken into account just on the throughput
via post-detection selection of the detected events. That this is
clearly not a valid inference could be seen from the situation of
B92, for which security is totally breached in an intercept-resend
attack when the loss is above a certain threshold determined by the
two signal states, or in any coherent-state BB84 protocol {[}17{]}.
\begin{figure}
\includegraphics[width=12cm]{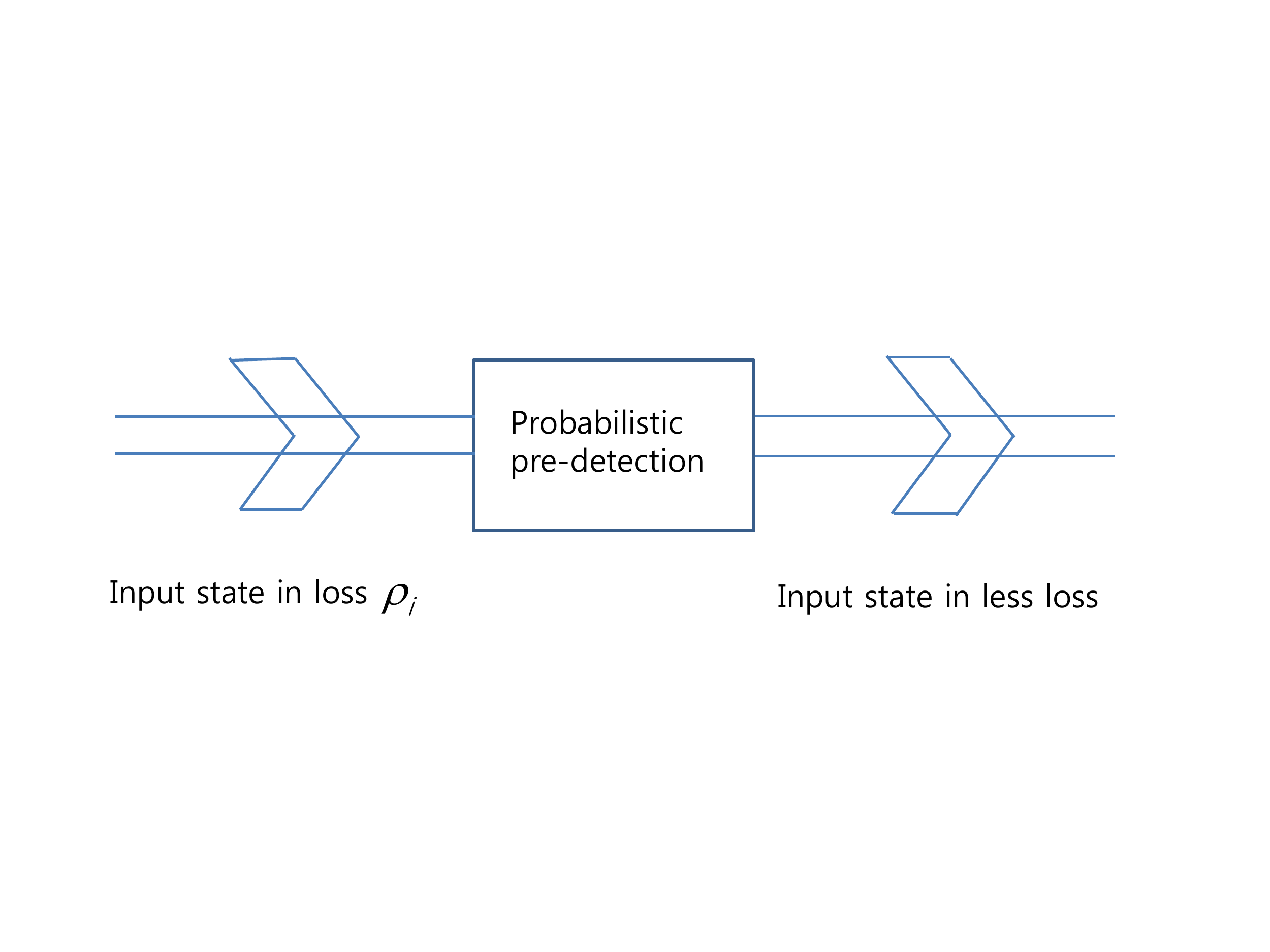}
\caption{\label{Fig-1} Schematic way to eliminate or reduce the effect of loss by user: loss is alleviated or eliminated with favorable pre-detection outcome.}
\end{figure}

\begin{figure}
\includegraphics[width=12cm]{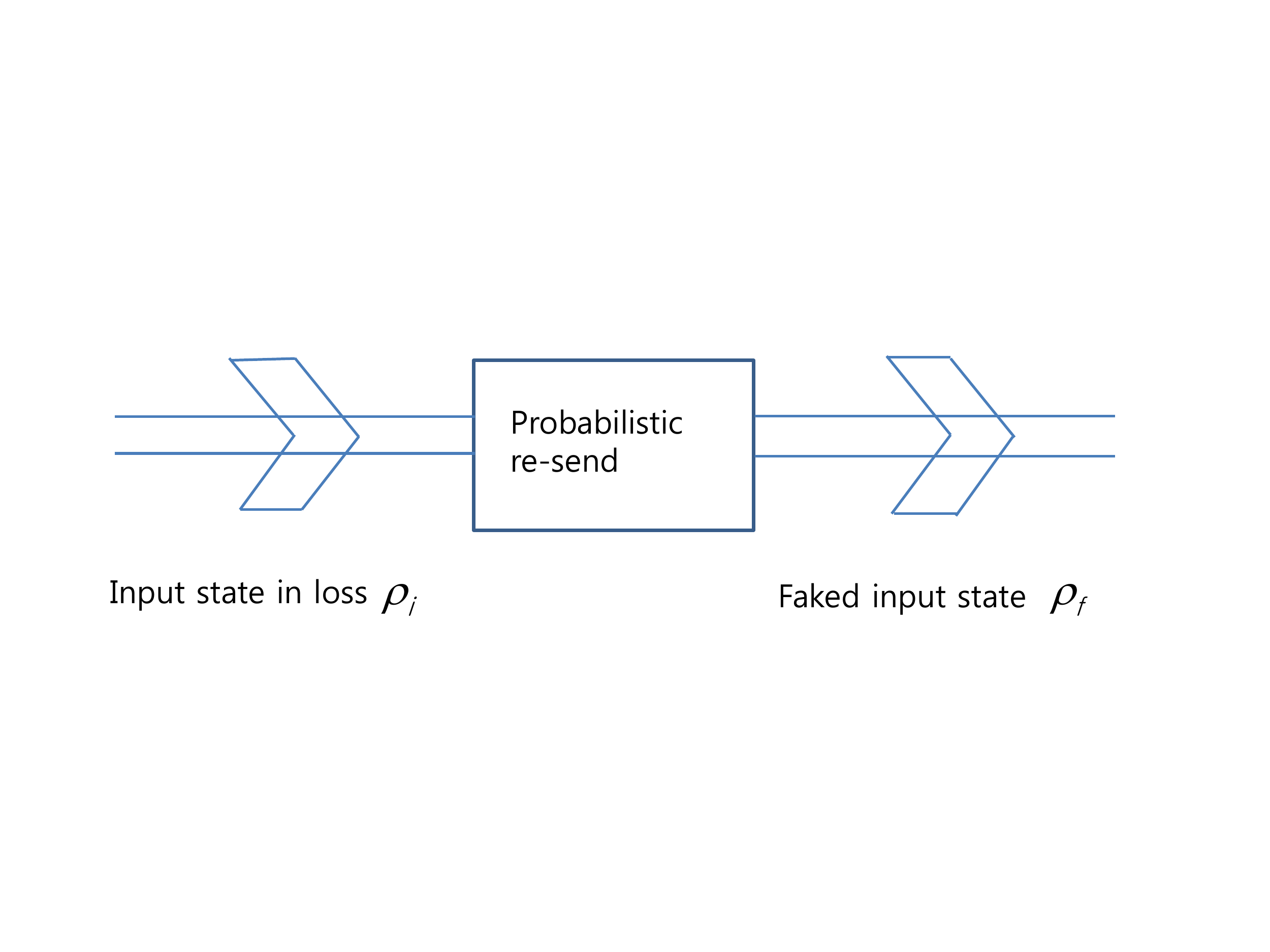}
\caption{Schematic way to take advantage of loss by attacker: a more favorable input state $\rho_f$ from Eve's viewpoint is sought with possible quantum signal detection (PRS attack).
\label{Fig-2}}
\end{figure}

Generally, the users may try to reduce loss by pre-detection as indicated
in Fig. 1, with success probability itself limited by the loss. Examples
include QND measurement and ``herald qubit amplifier''.
However, Eve also has a similar attack approach, the ``probabilistic
re-send (PRS) attack'' indicated in Fig. 2. Sufficient
loss would allow her to cover the deleted bits in principle and often
in practice also. PRS attacks include probabilistic approximate cloning
which is itself a generalization of the attack in ref {[}17{]} that
is equivalent to probabilistic exact cloning. Note that the possibility
of bit deletion from loss \textit{violates} the usual information-disturbance
tradeoff that underlines QKD security, in that information can be
gained by Eve without causing any relevant disturbance.

While PRS attacks can be covered in a sufficiently general formulation
on Eve's probe, it is not automatically covered by
merely bringing up the possible use of post-detection selection {[}18{]},
QND measurement or squashing {[}19-20{]}, or heralded qubit amplifier
{[}21{]}. Indeed, it does not seem a security proof covering all possible
PRS attacks in significant loss has ever appeared. The analysis of
ref {[}22{]} includes detector inefficiency but not transmission loss.
Absorbing transmission loss in the detector efficiency with $\rho_{i}$ replaced
by a state without loss is just the same as post-detection selection, in
addition to yielding a possibly very small detector efficiency. Note
that there is no complete security proof in loss even just under individual
attack. This grave consequence of loss on security has been pointed
out previously in {[}8{]}, and further elaborated in {[}23{]}.

\section{Problems of modeling versus side channel}
There are two kinds of mathematical modeling problems in QKD security
analysis of concrete systems:

\begin{enumerate}
\renewcommand{\theenumi}{(\Alph{enumi})}
\renewcommand{\labelenumi}{\theenumi}
\item whether the model includes typical general features of a real cryptosystem;
\item whether the operative assumptions of the security analysis are satisfied in the real system it is applied to.
\end{enumerate}

As an example of (A), the quantum signal state space in any QKD implementation
is never a qubit but an infinite-dimensional boson mode. That a different
dimension from two may breach security is clearly brought out in a
specific example {[}24{]}. The situation of loss is discussed above.
Thus, all qubit-based security proof is not directly applicable to
a real system, but such security claim was often made on the basis
of qubit proofs.

There are many examples of (B), such as the use of threshold detector
or Poisson source model for lasers without phase randomization. Much
more significant are the time-shift attack {[}25{]} based on detector
efficiency mismatch {[}26{]} and the blinding attacks {[}2-3,5{]}.
Detector efficiency mismatch has been dealt with in ref {[}18{]}.
What is unsettling about the time-shift attack is that the detailed
detection mechanism in the detector can be exploited to lead to a
huge mismatch. The blinding attacks (based on detector controllability
by Eve more essentially than ``faked state'')
is even more unsettling, because it does not lead to any common detector
imperfection representation and relies on the internal detector electronic
behavior. While the particular possibility of detector blinding can
be added in a security analysis {[}27{]}, it is not clear how one
would know all the relevant internal electronics behavior have been
included in any particular model. Some discussion on similar but more
general modeling question can be found in ref {[}28{]}. Note that
Fig. 2 can be used to represent timing and blinding attacks, when
the input state itself is already ``faked''
in a specific way by Eve and the detector electronic behavior and
total system loss may together allow an attack to succeed.

In this connection, it may be pointed out that this is not a ``side
channel'' issue as it is the case with the RSA timing
attack. Side channels can be closed once and forever, but the detector
in a QKD system is an integral part of the receiver one must have, part of the "main channel".
For example, if a detector leaks radiation of different characteristics
depending on the incoming state, it can be sealed and thus the leak is a side
channel. But the detection mechanism is not a side channel. Another
point is that a side channel would \textit{not} affect the original
cryptosystem representation, surely the case also for the RSA timing
attack. However, the system model has to be extended to include the
time-shift attack and blinding attack {[}27{]}. To what extent must
one model the internal behavior of a detector, or any system component,
so that the resulting security analysis captures all the relevant
features of the cryptosystem, instead of getting new surprises from
time to time? If this question is not settled there will be no security
proof for any concrete QKD system even just in principle, whatever
else one may have achieved. The detector representation problem goes
beyond (B) and squarely to (A) above, an issue of \textit{completeness
}of the cryptosystem model.

There are other systems that are not subject to such detector based
attacks, including continuous variable QKD which is, however, not
yet proved secure according to the standard view in the literature
{[}13{]}. The KCQ approach [6] is also immune to such attacks,
especially Y00 in any of the formats (PSK, ISK, QAM) that have been
studied, because there is essentially no deleted bit. However, general
IT security has not been established for any KCQ protocol.

\section{Outlook}
Security is a serious matter and cannot be established experimentally.
We see in the above that, just in principle, fundamental quantitative
security has not been properly addressed in QKD security analysis
and the effect of loss has not been properly accounted for. The widespread
perception of proven QKD security is based on several omission or
errors of reasoning:

\begin{enumerate}
\renewcommand{\theenumi}{(\roman{enumi})}
\renewcommand{\labelenumi}{\theenumi}
\item No proven security against known-plaintext attack when the generated
key $K$ is used for encryption;
\item Use of single-number constraint on the attacker\textquoteright{}s
probability distribution on $K$ when the number is not or cannot be
bounded tightly enough;
\item Not including all possible attacks in the presence of significant
transmission loss;
\item Not including relevant device characteristics.
\end{enumerate}

The situation is summarized in the following Table 1:
\begin{table}[!h]
%\begin{raggedright}
\begin{tabular}{|p{5cm}|p{6cm}|p{5cm}|}
\hline
& Perceived & Real
\\ \hline \hline
raw security of $K$ during key generation 
& criterion $d\le2^{-l}$ implies $K$ is perfect with probability $\ge1-\epsilon$ 
& probability Eve gets a subset $\tilde{K}\subseteq K$ can be as big as $2^{-l}$
\\ \hline
composition security of $K$ against known-plaintext attack when used in encryption 
& criterion $d\le2^{-l}$ implies $K$ is perfect withprobability $\ge1-\epsilon$ 
& $d$ is not the proper criterion
\\ \hline
privacy amplification 
& given Eve's entropy can make $d$  small and hence improve security 
& cannot improve $p_{1}$, Eve's maximum probability of getting the whole $K$
\\ \hline
key generation rate 
& $H(A|B)-H(A|E)$ 
& exponent $r$ of $p_{1}\sim2^{-rn}$
\\ \hline
determination of security
& by analysis of entropy hierarchy with respect to criterion $d$ 
& by bounding Eve's success probabilities
\\ \hline
effect of transmission loss on security 
& reduction of key rate but not security 
& many possible attacks by Eve not accounted for
\\ \hline
modeling of cryptosystem photon detector 
& a side channel issue 
& part of the completeness issue in original system representation
\\ \hline
\end{tabular}\caption{QKD Security Situation}
%\par\end{raggedright}
\end{table}

There are two ways to deal with these problems. The first is to limit
one's claim, for example to known-plaintext attacks with no quantum
memory. One can, say, wait an hour before using the generated key.
One can ignore joint attacks that require entanglement across modes
either in the probe or in the measurement. The resulting KPA security
appears provable for at least KCQ protocols and would still represents
major progress beyond what can be obtained with standard ciphers.
In any case it is better to avoid misleading terminology like ``unconditional'',
``no signaling'', and ``device
independent''. Generally, a more careful and critical
attitude in making security claim would be appropriate. The second
way is to look for new approaches or major modification of existing
ones. In particular, we need a general proof that device internal
electronics cannot lead to security loopholes in the protocol.

Note added for v.4: For new developments on the topics of sections II--IV, see [29]-[30].

\section*{Acknowledgements}
This work was supported by the Air Force Office of Scientific Research.


\begin{thebibliography}{27}
\bibitem{key-4}In this paper, examples of such alleged claims in
the literature would not be quoted. Also, by QKD we refer to protocols
involving intrusion level estimation which excludes KCQ protocols
{[}6{]}.
\end{thebibliography}
\end{document}